# Phonon Raman scattering of perovskite LaNiO$_3$ thin films

**N. Chaban, M. Weber, S. Pignard, J. Kreisel\***

Laboratoire des Matériaux et du Génie Physique, CNRS - Grenoble Institute of Technology, Minatec, 3, parvis Louis Néel, 38016 Grenoble, France

**Abstract**

We report an investigation of perovskite-type LaNiO$_3$ thin films by Raman scattering in both various scattering configurations and as a function of temperature. The room-temperature Raman spectra and the associated phonon mode assignment provide reference data for phonon calculations and for the use of Raman scattering for structural investigations of LaNiO$_3$, namely the effect of strain in thin films or heterostructures. The temperature-dependent Raman spectra from 80 to 900 K are characterized by the softening of the rotational $A_{1g}$ mode, which suggest a decreasing rhombohedral distortion towards the ideal cubic structure with increasing temperature.

\* to whom correspondence should be addressed (*E-mail: Jens.Kreisel@grenoble-inp.fr*)

Among ABO$_3$ perovskites, rare earth nickelates with the generic formula $RE$NiO$_3$ ($RE$ = Rare Earth) have attracted a lot of research interest since the report of a sharp metal-to-insulator (MI) transition whereof the critical temperature $T_{MI}$ can be tuned with the rare earth size. Further to this, most $RE$-nickelates exhibit in the insulating phase a complex anti-ferromagnetic ordering below the Néel temperature $T_N$ (see e.g. ref. [1,2]). A more recent interest in nickelates stems from their potential multiferroic properties[2-5].

Within the family of nickelate perovskites, LaNiO$_3$ (LNO) is the only nickelate which shows in its stoichiometric bulk form no temperature-induced MI transition but remains metallic down to the lowest temperature. This property makes LaNiO$_3$ one of the few possible electrodes in oxide-based electronics, namely in the field of strain-engineered perovskite thin films which attract increasing interest. Moreover, LNO recently attracts attention due to properties such as correlated Fermi liquid behavior[6], as oxygen sensor devices[7], predictions of superconductivity[8] or electric-field control of conductivity in LNO thin films[9]. Most of the recent work focuses on LNO thin films, while single-crystal growth of LaNiO$_3$ has remained unsuccessful until now. Finally, it has been shown that the intrinsic properties of LNO can be significantly modified through its incorporation in superlattices such as LaNiO$_3$/$A$MnO$_3$ (ref. [10,11]) or LaNiO$_3$/LaAlO$_3$ (ref.[12]), similarly to what is observed in other perovskite-type superlattices. In all these thin film or superlattice systems the structural characterization, namely the strain state between substrate and film and in superlattices, is essential for a better understanding of the observed physical properties.

In addition to the more commonly used X-ray diffraction and electron microscopy, Raman scattering has shown to be a versatile and non-destructive probe for investigating structural properties in thin oxide films. Applications of Raman spectroscopy, which probes zone centre phonons, range from more general aspects like subtle structural distortions in perovskites to the understanding of thin films characteristics such as strain effects, texture, X-ray amorphous phases, superlattice-related features, etc. [13-21] To the best of our knowledge there are neither literature reports on the experimental Raman signature of LaNiO$_3$ nor reports of phonon calculations. Even more generally, Raman investigations of perovskite nickelates are to date limited to NdNiO$_3$ (ref. [3,17]) and SmNiO$_3$ (ref. [3]).

Here we present an experimental Raman scattering study of LaNiO$_3$ thin films with the aim (*i*) to increase our current understanding of its phonons spectra, namely in relation with structural distortions (*ii*) to provide reference data for both future Raman investigations of bulk and thin film samples and future phonon calculations and (*iii*) to study its temperature-dependent behavior.

LaNiO$_3$ thin films have been deposited on Silicium (100) substrates by injection Metal-Organic Chemical Vapour Deposition (MOCVD) "band flash" using 2,2,6,6-tetramethylheptanedionato-chelates of corresponding metals as volatile precursors. Deposition conditions were 680°C using an atmosphere of argon-oxygen at 10 mbar pressure, followed by an in-situ annealing at ambient pressure under a pure oxygen flow. X-ray diffraction (XRD) pattern of the so-obtained films of 200 nm thickness illustrate well-crystallized LaNiO$_3$ films without traces of binary oxides such as La$_2$O$_3$ or NiO, within the diffraction detection limit. As expected, the deposited films on Si are polycrystalline.

Micro-Raman spectra were recorded using a LabRam Jobin-Yvon spectrometer with a spectral cut-off at ≈ 120 cm$^{-1}$. Our tests with three different wavelengths as exciting lines (488.0, 514.5 and 632.8 nm) illustrate that the overall spectral signature of the film is similar for the three wavelengths, as expected from Raman scattering without resonance effects. Similarly to what is commonly observed for metals, the spectrum obtained with the 632.8 nm laser line provides the best defined Raman spectra and thus this wavelength has been retained for our study. It is well-known that Raman spectra recorded on metallic oxides with a reduced laser penetration depth (i.e. *Re*-manganite[15,22], *Re*-nickelates[17] or La$_2$CuO$_4$ [23]) often show a strong dependence on the exciting laser power leading to structural modifications, phase transitions or even locally decomposed material. For LaNiO$_3$ we have observed that experiments with a laser power above several mW lead indeed to a modified spectral signature, characterized by a small but observable low-wavenumber shift of the LaNiO$_3$ modes illustrating a bond lengthening from local laser heating. As a consequence, our experiments have been carried out using laser powers of less than 1 mW under the microscope, and we have carefully verified that no structural transformations or overheating take place. Controlled temperature-dependent Raman measurements have been carried out by using a commercial LINKAM THS600 heating stage placed under the Raman microscope. After an initial cooling to liquid nitrogen temperature, the temperature-dependent spectra have been obtained by heating the sample from 80 to 900 K. The Raman spectra before and after temperature measurements are identical, attesting the reversibility of temperature-induced changes.

At ambient conditions LaNiO$_3$ crystallizes in a rhombohedrally distorted perovskite structure with space group $R\bar{3}c$. With respect to the ideal cubic $Pm\bar{3}m$ structure this rhombohedral structure is obtained by an anti-phase tilt of the adjacent NiO$_6$ octahedra about the $[111]_p$ pseudo-cubic diagonal, described by the $a^-a^-a^-$ tilt system in Glazer's notation[24]. The 10 atoms in the unit cell of rhombohedral LaNiO$_3$ give rise to 27 (**k** = 0) optical modes that can be characterized according to the space group using group theoretical methods. Such a treatment leads to only five

Raman-active modes $\Gamma_{Raman} = A_{1g} + 4\,E_g$. A vibrational pattern of all $A_{1g}$ and $E_g$ phonons in the $R\bar{3}c$ structure has been proposed by Abrashev *et al.* [25]. According to the Raman tensors of the $R\bar{3}c$ space group, a configuration of crossed polarizer's (*xy*-configuration) will lead to the observation of $E_g$ modes only, while parallel polarizer's (*xx*- or *yy*-configuration) allow the observation of both $A_{1g}$ and $E_g$ modes.

Figure 1.a presents a depolarized Raman spectrum of a polycrystalline $LaNiO_3$ film on a Si substrate. Besides a mode at 520 cm$^{-1}$ coming from the Si substrate, we observe 4 main Raman modes at 156, 209, 399 and 451 cm$^{-1}$. Taking into account that our experimental setup prohibits the observation of the rotational $E_g$ mode, which is expected below 100 cm$^{-1}$, this observation is in agreement with the group theoretical prediction for the $R\bar{3}c$ space group of $LaNiO_3$.

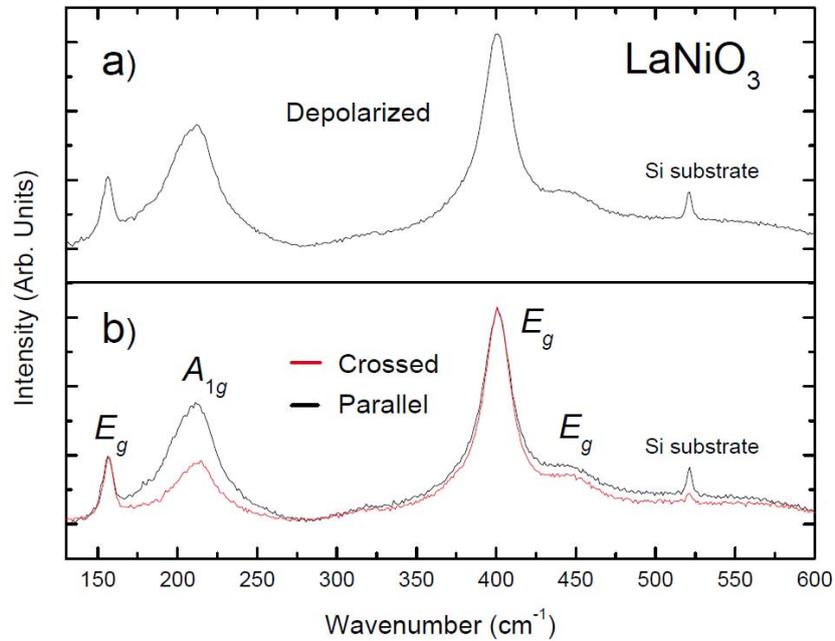

**Figure 1**  Depolarized (a) and polarized (b) Raman spectra of $LaNiO_3$ films on a Si substrate

Figure 1.b presents polarized Raman spectra of a $LaNiO_3$ film on Si. As expected from the Raman tensors, the spectrum with parallel polarizers is similar to the depolarized spectrum. On the other hand, the spectrum with crossed polarizers shows more pronounced changes, even though the textured nature of the investigated $LaNiO_3$ film precludes the observation of spectra with a clear extinction of Raman bands. The band at $\approx 209$ cm$^{-1}$ shows a significantly reduced intensity under crossed polarizer's and can thus be assigned to the single $A_{1g}$ phonon mode of LaNiO$_3$, while the modes at 156, 399 and 451 cm$^{-1}$ can be assigned to phonon modes of $E_g$ symmetry.

In the past, the empirical comparison of materials with a similar type of structure has been a rich source of understanding Raman signatures in oxide materials. For our study of LaNiO$_3$, the consideration of Raman data[25-28] for LaAlO$_3$, LaMnO$_3$ and LaCoO$_3$ is of specific interest, because they crystallize in the same $R\bar{3}c$ symmetry. We first compare the position in frequency of the $A_{1g}$ mode at 209 cm$^{-1}$ for LaNiO$_3$ to the positions observed for LaAlO$_3$, LaCoO$_3$ and LaMnO$_3$ at 132, 232 and 249 cm$^{-1}$, respectively[25-27]. The large difference in frequency between the three similar materials is at first sight surprising but can be understood by the fact that the $A_{1g}$ mode presents a rotational vibration pattern[25,26] and thus naturally scales with the rhombohedral distortion in terms of the octahedra tilt angle (i.e. the $A_{1g}$ mode is the soft mode driving the structural distortion in $R\bar{3}c$ perovskites[25,26]). An instructive illustration of this scaling is the behavior of the $A_{1g}$ mode in LaAlO$_3$ which softens to zero frequency at the rhombohedral-to-cubic phase transition which can be induced by increasing temperature[26] or pressure[29]. As a consequence, the different positions in frequency for the $A_{1g}$ mode in La$Me$O$_3$ perovskites point to a different and $Me$-dependent $Me$O$_6$ octahedra tilt angle, which is qualitatively in agreement with a tilt angle of for instance α = 5° for LaAlO$_3$ and α = 11° for LaMnO$_3$ (ref. [25]). By comparing the Raman spectra of an important number of perovskites Iliev et al.[30] has been going a step further by proposing that the room temperature frequency of rotational $A_{1g}$ modes scales with the tilt angle according to ≈ 23.5 cm$^{-1}$/deg. Although empirical in nature, Iliev's relation applies rather well for a number of materials[25,30-32]. In order to test this relationship for LaNiO$_3$, we have first calculated the NiO$_6$ octahedra tilt angle of α = 9° by using the reported[33] oxygen-site position in LaNiO$_3$ and the geometric relation[34] which links this position to the octahedra tilt angle. Using this tilt value and Iliev's relationship leads to an expected frequency of ≈ 211 cm$^{-1}$, which is very close to the experimentally observed value of 209 cm$^{-1}$ for the $A_{1g}$ mode in LaNiO$_3$. This very good agreement validates our assignment of this mode to a rotational $A_{1g}$ mode and suggests that this mode can be used in substituted LaNiO$_3$ or in strained LaNiO$_3$ thin films for probing changes in the octahedra rotation angle, which is one of the key parameters in the control of its electronic properties.

The comparison of LaNiO$_3$ with other La$Me$O$_3$ and the vibrational pattern described in literature[25] also allows assigning the $E_g$ modes: The mode at 156 cm$^{-1}$ corresponds to pure La vibrations in the hexagonal [001] plane along the $a$ and $b$-axis (at 163 cm$^{-1}$ in both LaAlO$_3$ and LaMnO$_3$ [25], and at 172 cm$^{-1}$ in LaCoO$_3$ [27]). This mode will be sensitive to substitutions on the $A$- site via mass changes. The high-frequency modes at 399 and 451 cm$^{-1}$ can be assigned to vibrational modes of the oxygen cage.

We now turn to the effect of temperature on the Raman-active phonons of LaNiO$_3$ and presents in Figure 2 Raman spectra from 80 K up to 900 K with sharp bands at low-temperature. With increasing temperature all bands show the expected thermal broadening, but the overall spectral signature is qualitatively maintained and no drastic changes occur, which suggests that LaNiO$_3$ maintains its $R\overline{3}c$ symmetry and undergoes no phase transition in the investigated temperature range. For a more quantitative analysis, Figure 3 illustrates the evolution of the band position as a function of temperature, as deduced from a spectral deconvolution. Note that the $A_{1g}$ mode could not be reliably determined above 550 K due to band broadening and overlap with the $E_g$ mode. Due to their different symmetry such a crossing is allowed.

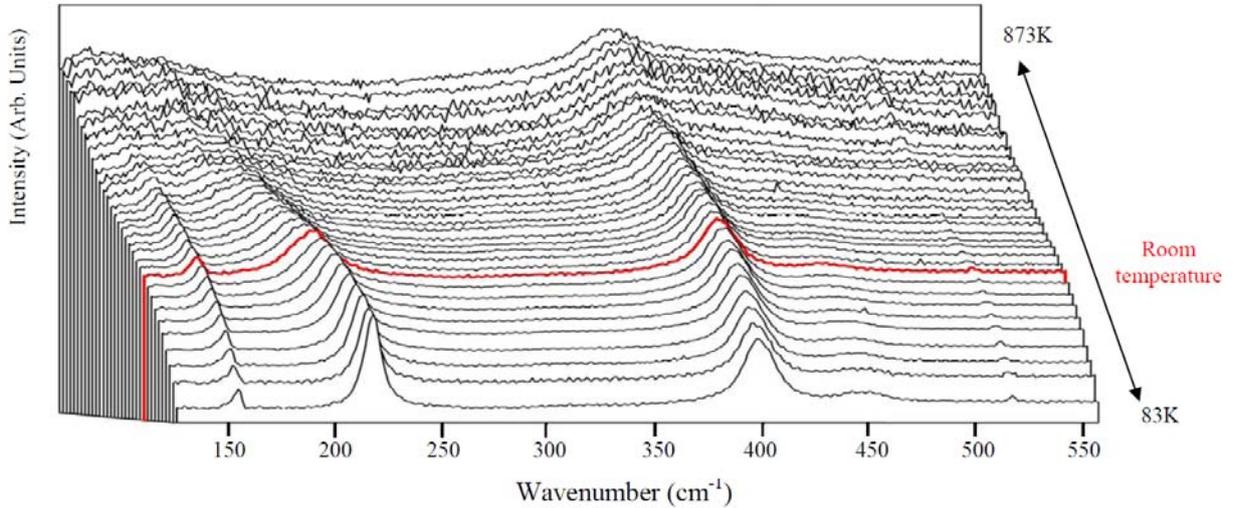

*Figure 2*

*Temperature-dependent Raman of a LaNiO$_3$ film from 83 K to 873 K, with a temperature increment of 50 K in-between the shown spectra.*

Generally speaking, all bands shift to lower frequencies with increasing temperature due to thermal expansion which conditions interatomic bond lengthening. However, the most eye-catching spectral signature with increasing temperature is the softening of the $A_{1g}$ mode from 222 cm$^{-1}$ at 80 K to 192 cm$^{-1}$ at 553 K. This pronounced shift provides further support for the assignment of this band to the rotational $A_{1g}$ soft mode of LaNiO$_3$, which is expected to soften stronger than other modes. As a consequence of this, the shift of the $A_{1g}$ soft mode in LaNiO$_6$ can be interpreted as a continuous decrease of the NiO$_6$ octahedra tilt angle with increasing temperature, similarly to what is observed in LaAlO$_3$ when approaching the temperature-induced rhombohedral-to-cubic phase transition [26,28].

From a phenomenological point of view, the behavior of a system in the vicinity of a symmetry-breaking structural phase transition can be described in the framework of Landau theory which was in the past successfully applied to the analysis of various tilted perovskite-type materials [35]. For the simplest second-order phase transition, the temperature dependence of the soft-mode frequency $\omega$ is in the low-symmetry phase ($T < T_c$) given by $\omega^2 = \omega_0^2(T_c-T)$. In the inset of Figure 3 we have plotted the temperature-dependent data of the $A_{1g}$-band showing straight-line behavior of $\omega^2$ vs. temperature thus validating that a Landau model can be used to describe the observed data and to evaluate via extrapolation a phase transition temperature of around 1780 K. Literature support for the viability of such an extrapolation even for data well below the phase transition comes from literature Raman work[26] on PrAlO$_3$ and NdAlO$_3$ where the soft modes permitted reasonable estimated extrapolations for phase transition temperatures.

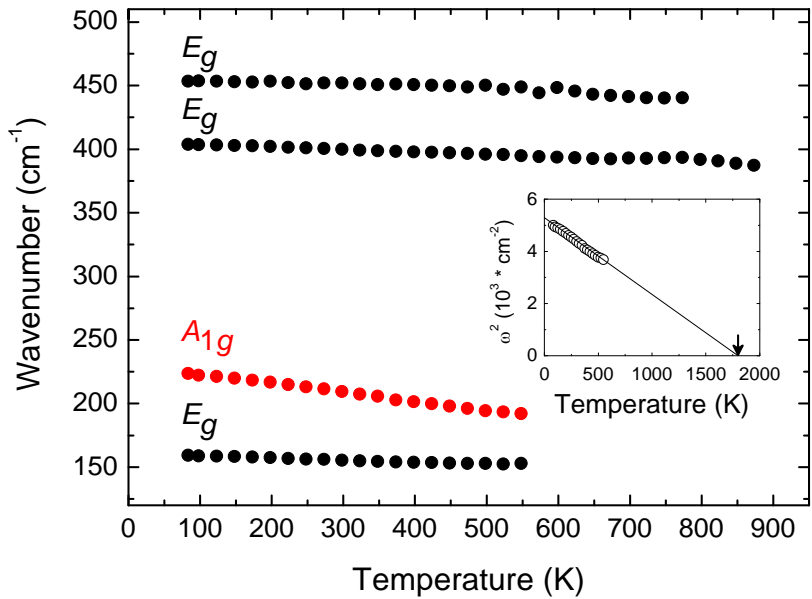

**Figure 3**

Temperature-dependent evolution of the $A_g$ and $E_g$ Raman mode positions in LaNiO$_3$. The inset shows the temperature-dependent evolution of the squared wavenumber for the soft rotational $A_{1g}$ band. The solid curve corresponds to least-squares fits according to $\omega^2 = \omega_0^2(T_c-T)$ with $\omega_0 =$ 229.4 cm$^{-1}$ and $T_c =$ 1780 K;

We have presented a Raman scattering investigation of LaNiO$_3$ thin films which allowed to observe four out of the five Raman-active phonon modes. A symmetry assignment has been proposed on the basis of the analysis of various polarisation configurations. The presented experimental reference data set the basis for phonon calculations and for the use of Raman scattering in LaNiO$_3$ samples, for instance for the investigation of strain (via phonon shifts) in thin films and heterostructures.

Furthermore, we have presented a temperature-dependent Raman scattering investigation of LaNiO$_3$ from 80 K up to 900K. Within this temperature LaNiO$_3$ undergoes no structural phase transition since no drastic changes like band splitting or band appearance/disappearance are observed. However, the pronounced softening of the rotational $A_{1g}$ mode points at a decreasing rhombohedral distortion towards the ideal cubic perovskite structure, which we expect from our extrapolation to occur at $\approx$ 1800K.

**Acknowledegement**

The authors acknowledge financial support from a "BQR grant" of the Grenoble Institute of Technology.